\documentclass[onecolumn,floatfix,showpacs,showkeywords,10pt,nofootinbib]{revtex4}
\pdfoutput=1
\usepackage{amssymb,amsmath}
\usepackage{graphicx,float}
\usepackage{indentfirst}
\usepackage{indentfirst}
\newcommand{\be}{\begin{equation}}   
\newcommand{\ee}{\end{equation}}

\newcommand{\ba}{\begin{eqnarray}}
\newcommand{\ea}{\end{eqnarray}}

\newcommand{\lie}{\pounds_{\bf n}}

\usepackage{float}

\newcommand{\bege}{\begin{equation}}
\newcommand{\bpartial}{\mathop{\partial\kern -4pt\raisebox{.8pt}{$|$}}}
\newcommand{\enge}{\end{equation}}
\newcommand{\beq}{\begin{eqnarray}}
\newcommand{\benu}{\begin{enumerate}}

\newcommand{\enu}{\end{enumerate}}
\newcommand{\eeq}{\end{eqnarray}}

\newcommand{\noi}{\noindent}

\begin{document}

\title{Casadio-Fabbri-Mazzacurati Black Strings and Braneworld-induced Quasars Luminosity Corrections}

\author{Rold\~ao da Rocha}
\email{roldao.rocha@ufabc.edu.br} \affiliation{Centro de
Matem\'atica, Computa\c c\~ao e Cogni\c c\~ao, Universidade
Federal do ABC 09210-170, Santo Andr\'e, SP, Brazil.}
\author{A. Piloyan}
\email{arpine.piloyan@ysu.am}
\affiliation{Institut f\"ur Theoretische Physik, Philosophenweg 16
D-6912, Heidelberg, Germany\\Yerevan State Univ., Faculty of Physics, Alex Manoogian 1, Yerevan 0025, Armenia}
\author{A. M. Kuerten}
\email{andre.kuerten@ufabc.edu.br} \affiliation{Centro de
Ci\^encias Naturais e Humanas, Universidade
Federal do ABC 09210-170, Santo Andr\'e, SP, Brazil.}

\author{C. H. Coimbra-Ara\'ujo}
\email{carlos.coimbra@ufpr.br}
\affiliation{Campus Palotina,
Universidade Federal do Paran\'a, UFPR, 85950-000, Palotina, PR,
Brazil.}

\begin{abstract}{
This paper aims to  evince  
the corrections on the black string warped horizon in the braneworld paradigm, and their drastic physical consequences, as well as  to provide subsequent applications in astrophysics. Our analysis concerning black holes on the brane  departs from the Schwarzschild case, where the black string is unstable to large-scale perturbation. The cognizable measurability of the black string horizon corrections due to braneworld effects
is investigated, as well as their applications in the variation of quasars luminosity.   
We delve into the case wherein 
 two solutions of Einstein's equations proposed by Casadio, Fabbri, and Mazzacurati, regarding black hole metrics presenting a post-Newtonian
parameter measured on the brane. In this scenario, it is possible to analyze purely the braneworld corrected variation in quasars luminosity, by an appropriate  choice of the post-Newtonian parameter that precludes Hawking radiation on the brane: the variation in quasars luminosity is uniquely provided by pure braneworld effects, as the Hawking radiation on the brane is suppressed.
 }
\end{abstract}

\maketitle

\flushbottom
\section{Introduction}
Black holes solutions of Einstein equations in general relativity are useful tools to investigate the space-time structure and  underlying models for gravity and its quantum effects, as well as to study the astrophysics regarding supermassive objects, for instance.
Extra-dimensional space-times are scenarios for extensions of the general relativity, providing solutions to the Einstein's field equations, as   black holes metrics in higher dimensions, and some ensuing applications to cosmology in such a context. In addition,  the recent effort to deal with the
hierarchy problem, by inducing gravity to leak into extra dimensions \cite{todos}, is explored in braneworld models. They are based on M-theory and string theory   \cite{PREV,hw, zwi}. In particular, an useful approach to deal with the hierarchy is provided an effective 5D reduction of the Ho$\check{\rm r}$ava-Witten theory
\cite{Randall2,hw,luka1}.

Impelled by a thorough development concerning  gravity on 5D braneworld scenarios and some applications, in particular the black holes/black strings horizons variations induced by braneworld effects,  \cite{nossos,adsbranes,nossoprd,ROLDAO/CARLAO,ruth33,meuhoff}, further aspects concerning corrections in the black string like objects and their warped horizons are here introduced. 
The Casadio-Fabbri-Mazzacurati metrics on the brane, namely the type I and type II  black hole solutions  \cite{rs06,rs09} are now analyzed and regarded as generating the bulk metric, inducing a black string  like warped horizon. This
procedure is well known for the  Schwarzschild metric \cite{nossoprd,ROLDAO/CARLAO,meuhoff,Maartens}. The  Casadio-Fabbri-Mazzacurati metrics depart from the Schwarzschild solution, possessing a post-Newtonian parameter. For some particular choice of this parameter, the black hole Hawking radiation on the brane is suppressed  \cite{rs06,rs09}. The black holes Hawking radiation  in braneworld scenarios was comprehensively investigated in, e. g.,  \cite{Casadioharms, card1}.

This article  is organized as follows: in Sec. \ref{2}, after presenting the Einstein field equations in the brane, the deviation in Newton's 4D gravitational potential is revisited.  For a static spherical metric 
on the brane, the propagating effect of 5D gravity is
evinced from the Taylor expansion (along the extra dimension) of the metric. Such expansion is accomplished in powers of the normal coordinate --- out of the brane --- which  provides the black string warped horizon profile.
Such expansion can  provide the bulk metric uniquely from 
the metric on the brane.
In Sec. \ref{3}, the type I and type II Casadio-Fabbri-Mazzacurati black string solutions and their respective warped horizons are obtained, analyzed and depicted. We analyze such solutions in the particular case where the associated post-Newtonian parameter makes the black hole Hawking   radiation to be suppressed. Such analysis has paramount importance, since 
 to measure pure effects to the corrections (by braneworld effects) for quasars luminosity is aimed. 
In Sec. \ref{4}, for an illustrative model for accretion in a supermassive black hole, the variation of luminosity in quasars 
is investigated more precisely for the two models provided by Casadio-Fabbri-Mazzacurati, and compared to the pure Schwarzschild black string. The correction effects on the black string warped horizon, induced and generated by braneworld models, preclude the Hawking radiation on the brane for the above mentioned suitable choice of the post-Newtonian  parameter. 
All results are illustrated by graphics 
and figures, and the quasars luminosity provided by the Casadio-Fabbri-Mazzacurati black hole solution is compared with the Schwarzschild one.

\section{Black string behavior along the extra dimension}
\label{2}
Hereupon the notation in \cite{Maartens,GCGR,Gergely:2001tn}
 is adopted, where $\{\theta_\mu\}$, {\footnotesize{$\mu = 0,1,2,3$}}, typify a basis for the cotangent space $T^\ast_xM$ at a point $x$ on a brane
$M$ embedded in a bulk.
A frame $\theta^A = dx^A$ ({\footnotesize{$A=0,1,2,3,4$}}) in the bulk is represented in local coordinates.  In  the brane defined by $y=0$,  [hereon $y$ denotes the associated Gaussian coordinate]  $dy = n_Adx^A$ is orthogonal to the brane. 
 The 
metric $\mathring{g}_{AB}dx^A dx^B = g_{\mu\nu}(x^\alpha,y)\,dx^\mu dx^\nu + dy^2$ endows the bulk, which is related to the brane metric $g_{\mu\nu}$ by 
$
 g_{\mu\nu} = \mathring{g}_{\mu\nu} - n_\mu n_\nu.
$ The bulk indexes $A,B = 0,\ldots,3$, as $\mathring{g}_{44} = 1$ and $\mathring{g}_{\mu 4} = 0$  \cite{Maartens}.
Hereupon the standard relations $\Lambda_4=\frac{\kappa_5^{2}}{2}\Big(\frac{1}{6}\kappa_5^{2}\lambda^{2}+\Lambda\Big)$ and $\kappa^2_{4}=\frac{1}{6}\lambda\kappa^4_5$ are considered, 
where $\Lambda_4$ denotes the effective brane cosmological constant, and $\lambda$ is
the brane tension. The constant  $\kappa_5 = 8\pi G_5$, where 
$G_5$ is the 5D Newton gravitational constant, denotes the 5D gravitational coupling, related to the
4D gravitational constant $G$ by $G_5 = G\ell_{\rm Planck}$, where $\ell_{\rm Planck} = \sqrt{G\hbar/c^3}$ is the Planck length.  The junction condition provides the extrinsic curvature tensor $K_{\mu\nu} = \frac{1}{2} {\lie} g_{\mu\nu}$ by  \cite{is,Maartens}
 \be
K_{\mu\nu}=-\frac{1}{2}\kappa_5^2 \left(T_{\mu\nu}+ \frac{1}{3}
\left(\lambda-T\right)g_{\mu\nu} \right),
 \label{junction}\ee
where $T=T^\mu{}_\mu $ is the trace of the energy-momentum tensor.  
The 5D Weyl tensor is given by
$
 C_{\mu\nu\sigma\rho} = {}^{(5)}R_{\mu\nu\sigma\rho} - \frac{2}{3} (\mathring{g}_{[\mu\sigma} {}^{(5)}R_{\nu]\rho} + \mathring{g}_{[\nu\rho} {}^{(5)}R_{\mu]\sigma}) - \frac{1}{6} {}^{(5)}R ( \mathring{g}_{\mu[\sigma} \mathring{g}_{\nu\rho]}),$ where ${}^{(5)}R_{\mu\nu\sigma\rho}$ denotes the components of the bulk Riemann tensor (as usual ${}^{(5)}R_{\mu\nu}$ and ${}^{(5)}R$ are the associated Ricci tensor and the scalar curvature). The trace-free and symmetric  components, respectively denoted by   ${\cal B}_{\mu\nu\alpha} = g_\mu^{\;\rho} g_\nu^{\;\sigma}
C_{\rho\sigma\alpha\beta}n^\beta$ and ${\cal E}_{\mu\nu} = C_{\mu\nu\sigma\rho} n^\sigma n^\rho$,  denote the  magnetic and electric  Weyl tensor components, respectively. \subsection{Brane field equations}

The Einstein brane field equations can be expressed as
\begin{eqnarray}\label{123}
&&G_{\mu\nu} =  - {\cal E}_{\mu\nu} -\frac{1}{2}{\Lambda}_5g_{\mu\nu}
+ \frac{1}{4}\kappa_5^4\left(\frac{1}{2}g_{\mu\nu}\left(T^2 - T_{\alpha\beta}T^{\alpha\beta}\right)+TT_{\mu\nu} - T_{\mu \alpha} T^\alpha _{ \;\,\nu}\right).\nonumber
\end{eqnarray}
\noindent The Weyl tensor electric term ${\cal E}_{\mu\nu}$  carries an imprint of high-energy effects sourcing Kaluza-Klein (KK) modes. The gravitational potential $V(r) =\frac{GM}{c^2r}$, associated to the 4D classical gravity,  is corrected by extra-dimensional effects \cite{Maartens,Randall2}:
\begin{equation}\label{potential1}
V(r) = \frac{GM}{c^2r}\left(1 + \frac{2\ell^2}{3r^2} + \cdots\right).
\end{equation}
\noindent
The parameter $\ell$ is associated with the bulk curvature radius and corresponds
to the effective size of the extra dimension probed by a 5D graviton \cite{Likken, Randall2,Maartens}.
Indeed, the contribution of the massive KK modes sums to a correction of the 4D  potential. At small scales $r \ll  \ell$, one obtains the 5D features related to the potential $V (r) \approx GM\ell/r^2$. For $r\gg \ell$ the potential is provided by (\ref{potential1})  reinforcing the gravitational field  \cite{Maartens,Randall2,ROLDAO/CARLAO}.

Considering vacuum on the brane, where $T_{\mu\nu} = 0$ outside a black hole, the field equations $G_{\mu\nu} =  - {\cal E}_{\mu\nu} - \frac{1}{2}\Lambda_ 5g_{\mu\nu}$ and $R = 0 = {\cal E}^{\mu}_{\;\; \mu}$ hold for braneworlds with $\mathbb{Z}_2$-symmetry. The vacuum field equations in the brane are 
${\cal E}_{\mu\nu} = - R_{\mu\nu},$ 
where the bulk cosmological constant is comprised into the warp factor. The bulk can host for instance a plethora of non standard model fields, as dilatonic or moduli fields \cite{Gergely:2006hd}.
Although a Taylor expansion of the metric  was used to probe properties of a black hole on the brane in, e. g., \cite{Dadhich,Maartens}, in order to enhance the range of our analysis throughout this paper 
a more complete approach to analyze braneworld corrections in the black string profile can be accomplished, based on \cite{meuhoff}.  

A Taylor expansion of the metric along the extra dimension allows us to analyze the black string more deeply.
The effective field equations are complemented by other ones, obtained from the 5D Einstein and Bianchi equations in Refs. \cite{GCGR,Maartens,Gergely:2001tn}. Hereupon, since we are concerned with the Taylor expansion of the metric along the extra dimension up to the fourth order, besides the effective field equation ${\bf \pounds}_{\bf n} K_{\mu\nu}= K_{\mu\alpha}K^\alpha
{}_\nu - {\cal E}_{\mu\nu}-\frac{1}{6}\Lambda_5 g_{\mu\nu}$, the  effective equations are considered:
\begin{eqnarray}
{\bf \pounds}_{\bf n} {\cal E}_{\mu\nu}  &=& \nabla^\alpha
{\cal B}_{\alpha(\mu\nu)} + \left(K_{\mu\alpha}K_{\nu\beta}
-K_{\alpha\beta}K_{\mu\nu}\right)K^{\alpha\beta}
+K^{\alpha\beta}R_{\mu\alpha\nu\beta}-K{\cal E}_{\mu\nu}+\frac{1}{6}
\Lambda_5\left(K_{\mu\nu}-g_{\mu\nu}K\right) \nonumber\\&&
\qquad\qquad+3K^\alpha{}_{(\mu}{\cal
E}_{\nu)\alpha}, \nonumber
\\  {\bf \pounds}_{\bf n} {\cal B}_{\mu\nu\alpha}&=&K_\alpha{}^\beta {\cal
B}_{\mu\nu\beta} -2\nabla_{[\mu}{\cal E}_{\nu]\alpha}-2{\cal B}_{\alpha\beta [\mu }K_{\nu]}{}^\beta.
\nonumber
\end{eqnarray}
These expressions are used to compute the terms in the Taylor expansion of the metric, along the extra dimension, providing the black string profile and further physical consequences as well. 
The effective field equations above were employed to construct a covariant analysis of the weak field~\cite{GCGR}. Denoting $K = K_\mu^{\;\,\mu}$, the Taylor expansion is given by \cite{meuhoff} [hereon we denote $g_{\mu\nu}(x,0) = g_{\mu\nu}$]:
 \ba
&&\hspace*{-0.6cm}g_{\mu\nu}(x,y)= g_{\mu\nu}(x,0)-\kappa_5^2\left[
T_{\mu\nu}+\frac{1}{3}(\lambda-T)g_{\mu\nu}\right]\,|y|+ \nonumber\\
&&~~{}+\left[-{\cal E}_{\mu\nu} +\frac{1}{4}\kappa_5^4\left(
T_{\mu\alpha}T^\alpha{}_\nu +\frac{2}{3} (\lambda-T)T_{\mu\nu}
\right) +\frac{1}{6}\left( \frac{1}{6}
\kappa_5^4(\lambda-T)^2-\Lambda_5
\right)g_{\mu\nu}\right]\, y^2+ \nonumber\\
&& +\left.\Bigg[2K_{\mu\beta}K^{\beta}_{\;\,\alpha}K^{\alpha}_{\;\,\nu} - ({\cal E}_{\mu\alpha}K^{\alpha}_{\;\,\nu}+K_{\mu\alpha}{\cal E}^{\alpha}_{\;\,\nu})-\frac{1}{3}\Lambda_5K_{\mu\nu}-\nabla^\alpha{\cal B}_{\alpha(\mu\nu)} + \frac{1}{6}
\Lambda_5\left(K_{\mu\nu}-g_{\mu\nu}K\right)
 \right.\nonumber\\
&&\hspace*{-0.4cm}\left.+K^{\alpha\beta}R_{\mu\alpha\nu\beta}+3K^\alpha{}_{(\mu}{\cal
E}_{\nu)\alpha}-K{\cal E}_{\mu\nu}+\left(K_{\mu\alpha}K_{\nu\beta}
-K_{\alpha\beta}K_{\mu\nu}\right)K^{\alpha\beta}-\frac{\Lambda_5}{3}K_{\mu\nu}\Bigg]\;\frac{|y|^3}{3!} +  \right.\nonumber\\&&+\left.\Bigg[\frac{\Lambda_5}{6}\left(R-\frac{\Lambda_5}{3} + K^2\right)g_{\mu\nu} + \left(\frac{K^2}{3}- \Lambda_5\right)K_{\mu\alpha}K^{\alpha}_{\;\,\nu} + \left(3\,K^\alpha_{\;\,\mu}K^{\beta}_{\;\,\alpha}-K_{\mu\alpha}K^{\alpha\beta}\right){\cal E}_{\nu\beta}\right.\nonumber\\&+&\left. \left(K^{\alpha}_{\;\,\sigma}K^{\sigma\beta} + {\cal E}^{\alpha\beta} +KK^{\alpha\beta}\right)\,R_{\mu\alpha\nu\beta} - \frac{1}{6}\Lambda_5R_{\mu\nu}
 + 2 K_{\mu\beta}K^{\beta}_{\;\,\sigma}K^\sigma_{\;\,\alpha}K^\alpha_{\;\,\nu} + K_{\sigma\rho}K^{\sigma\rho}K\,K_{\mu\nu}\right.\nonumber\\&+&\left.
  {\cal E}_{\mu\alpha}\left(K_{\nu\beta}K^{\alpha\beta}-3K^\alpha_{\;\,\sigma}K^{\sigma}_{\;\,\nu} + \frac{1}{2}KK^\alpha_{\;\,\nu}\right)+\left(\frac{7}{2}KK^\alpha_{\;\,\mu}- 3K^\alpha_{\;\,\sigma}K^\sigma_{\;\,\mu}\right){\cal E}_{\nu\alpha}-\frac{13}{2}K_{\mu\beta}{\cal E}^\beta_{\;\,\alpha}K^{\alpha}_{\;\,\nu} \right.\nonumber\\&+&\hspace*{-0.1cm}\left. (R-\Lambda_5 + 2K^2){\cal E}_{\mu\nu} - K_{\mu\alpha}K_{\nu\beta}{\cal E}^{\alpha\beta} - \frac{7}{6}K^{\sigma\beta}K^{\;\,\alpha}_{\mu}R_{\nu\sigma\alpha\beta} - 4K^{\alpha\beta}R_{\mu\nu\gamma\alpha}K^{\gamma}_{\;\beta}\Bigg]\,\frac{y^4}{4!} + \cdots \right.\label{tay} \ea
Such an expansion was analyzed in \cite{Maartens, c111} only up to the second order, although it fizzled out to explain more reliably the black string horizon behavior along the extra dimension.  In addition, this higher order expansion  provides further physical features regarding variable tension braneworld scenarios, since the expansion terms beyond second order  provide drastic modifications in the stability of black strings \cite{meuhoff}. For an alternative method which does not take into account the $\mathbb{Z}_2$-symmetry,  and some subsequent applications,  see \cite{Jennings:2004wz}. 

 For a vacuum in the brane,  Eq.(\ref{tay}) reads
 \beq
\hspace*{-.8cm}g_{\mu\nu}(x,y)&=&g_{\mu\nu}-\frac{1}{3}\kappa_5^2\lambda g_{\mu\nu}\,|y|~~{}+\left[\frac{1}{6}\left(\frac{1}{6}\kappa_5^4\lambda^2 - \Lambda_5\right)g_{\mu\nu}-{\cal E}_{\mu\nu} \right]\, y^2\nonumber\\&&-\frac{1}{6}\left(\left(\frac{193}{36}\lambda^3\kappa_5^6 +\frac{5}{3}\Lambda_5\kappa_5^2\lambda\right)g_{\mu\nu}+\kappa_5^2{R}_{\mu\nu}\right)\,\frac{|y|^3}{3!} + \nonumber\\&&+\left.\Bigg[\frac{1}{6}\Lambda_5\left(\left(R-\frac{1}{3}\Lambda_5 - \frac{1}{18}\lambda^2\kappa_5^4\right)+\frac{7}{324}\lambda^4\kappa_5^8\right)g_{\mu\nu} + \left(R -\Lambda_5 + \frac{19}{36}\lambda^2\kappa_5^4\right){\cal E}_{\mu\nu}\right.\nonumber\\
&&\left. \qquad + \frac{1}{6}\left(\frac{37}{36}\lambda^2\kappa_5^4- \Lambda_5\right)R_{\mu\nu}+ {\cal E}^{\alpha\beta}\,R_{\mu\alpha\nu\beta}\Bigg]\,\frac{y^4}{4!} + \cdots \right.\label{expandir}
\eeq
This expression is shown to be prominently relevant for our subsequent analysis. 

{ Hereon, the black hole horizon evolution along the extra dimension --- the warped horizon \cite{clark} --- shall be investigated, exploring the component $g_{\theta\theta}(x,y)$ in (\ref{expandir}).
Indeed, let us consider any spherically symmetric metric associated to a black hole --- in particular the Schwarzschild and the Casadio-Fabbri-Mazzacurati ones here investigated. Such metric has the radial coordinate given by $\sqrt{g_{\theta\theta}(x,0)} = r$. 
The black hole solution, namely, the black string solution \emph{on the brane}, is regarded when 
$\sqrt{g_{\theta\theta}(x,0)} = R$, where $R$ denotes the coordinate singularity, usually calculated by the
component $g_{rr}^{-1} = 0$ in the metric\footnote{Such calculation was also considered in Eq.(II.6) of \cite{Casadioharms} in the braneworld context.}. In the Schwarzschild metric $R = R_S = \frac{2GM}{c^2r}$.
 The coordinate singularities for the Casadio-Fabbri-Mazzacurati metrics are going to be analyzed in what follows, in the black string context as well. Such singularities shall be shown to be also  physical singularities (associated to the black holes and the black strings as well), by analyzing their respective four- and 5D Kretschmann scalars.
 In other words, in the analysis regarding the black string behavior along the extra dimension, we are concerned merely about the warped horizon behavior, which is provided uniquely by the value for the metric on the brane $\sqrt{g_{\theta\theta}(x,y)}\vert_{r = R_S}$.  More specifically, the black string
 horizon for the Schwarzschild metric ---  or warped horizon \cite{clark} --- is defined when the radial coordinate $r$ has the value $r = R_S =  \frac{2GM}{c^2}$, which is obtained when the 
  coefficient $\left(1-\frac{2GM}{c^2r}\right) = g_{rr}$ of the term $dr^2$ in the metric goes to infinity \cite{Casadioharms}. It corresponds to the black hole horizon on the brane. On the another hand, the (squared) general radial coordinate in spherical coordinates legitimately appears as the term $g_{\theta\theta}d\theta^2 = r^2 d\theta^2$ in the Schwarzschild metric. 
 Our analysis of the term $g_{\theta\theta}(x,y)$ (given by Eq.(\ref{tay}) for {\footnotesize{$\mu=\theta=\nu$}} as the most general case, and 
 provided by Eq.(\ref{expandir}) for the Schwarzschild metric) holds for any value $r$. In particular, the term originally coined ``black string'' 
 corresponds to the Schwarzschild case \cite{clark}, defined by the black hole horizon evolution along the extra dimension into the bulk. Hence, the black string  regards solely the so called ``warped horizon'', which is $g_{\theta\theta}(x,y)$, for the particular case where $r = R_S$ is a coordinate singularity. }

\section{Casadio-Fabbri-Mazzacurati braneworld solutions}
\label{3}

The analysis of the gravitational field equations on the brane is not straightforward, due to the fact that the propagation of gravity into the bulk does not allow a complete presentation of the brane gravitational field equations as a closed form system \cite{GCGR}. The investigation concerning the gravitational collapse on the brane is therefore very complicated \cite{caliper}. The solutions provided by Casadio, Fabbri, and Mazzacurati for the brane black holes metrics \cite{Dadhich,maart01,rs06} 
take into account the post-Newtonian
parameter $\beta$, measured on the brane. The case $\beta = 1$ generates forthwith an exact
Schwarzschild solution on the brane, and elicits a black string prototype. Furthermore, it was observed in \cite{rs06,rs09} that $\beta \approx 1$ holds in solar system scales
measurements \cite{c111}. The parameter $\beta$ is, furthermore, capable to indicate and to measure the difference between
the inertial mass and the gravitational mass of a
test body. This parameter also affects the perihelion shift and provides the Nordtvedt effect \cite{c111}.  Moreover, measuring $\beta$ gives information
about the vacuum energy of the braneworld or,
equivalently, the cosmological constant  \cite{rs06,rs08,rs09}. 

One of the main motivation regarding the Casadio-Fabbri-Mazzacurati setup is that black
holes solutions of the Einstein equations on the brane must depart from the Schwarzschild solution. In particular, the Schwarzschild associated black string is unstable to large-scale perturbations \cite{hawk,gre}: the associated Kretschmann scalar, regarding the 5D curvature, 
diverges on the Cauchy horizon \cite{rs06,Maartens}. 
{ 
Indeed, it is important to emphasize that, as we shall see for the Casadio-Fabbri-Mazzacurati black string, it might be possible to find out points along the extra dimension for which  the Kretschmann scalar ${}^{(5)}K =  {}^{(5)}R_{\mu\nu\rho\sigma} {}^{(5)}R^{\mu\nu\rho\sigma}$ diverges, i. e., they are indeed naked singularities along the extra dimension. For instance, in order to identify singularities, for a Schwarzschild black string  ${}^{(5)}K \propto 1/r^6$ \cite{hawk}. Hence there is a line singularity at $r=0$ along the extra dimension, but not at the Schwarzschild horizon  \cite{Maartens, clark}. Since the pure black string configuration is unstable \cite{gre}, this structure is not physical ab initio. Anyway for $y=0$ one reproduces the Kretschmann scalars standard 4D behavior. }
For the
Schwarzschild solution, the singularity on the brane extends
into the bulk and makes the AdS horizon singular. 
The Casadio-Fabbri-Mazzacurati black string solutions and their respective braneworld corrections are going to be presented and their stability analyzed as well.
For the sake of completeness the next Section is briefly devoted to the braneworld corrections to the Schwarzschild
solution \cite{Maartens,nossos}.

A static spherical metric on the brane is provided by $
g_{\mu\nu}dx^{\mu}dx^{\nu} = - F(r)dt^2 + ({H(r)})^{-1}{dr^2} + r^2d\Omega^2$. The Schwarzschild metric is corresponds to  $F(r)=H(r)=1-\frac{2GM}{c^2r}$.
Obtaining such functions remains an open problem in the black hole theory on the brane \cite{Maartens,
rs06,rs09}.
Considering the Weyl tensor projected electric component on the brane $
 {\cal E}_{\theta\theta} =0$ \cite{Maartens}  yields \cite{meuhoff}
\begin{eqnarray}
g_{\theta\theta}(r,y)&=&r^2\left[1-\frac{\kappa_5^2\lambda}{3}\,|y|~~{}+\frac{1}{6}\left(\frac{1}{6}\kappa_5^4\lambda^2 - \Lambda_5\right)\, y^2-\left(\frac{193}{216}\lambda^3\kappa_5^6 +\frac{5}{18}\Lambda_5\kappa_5^2\lambda\right)\,\frac{|y|^3}{3!} + \right.\nonumber\\&&\qquad\qquad\qquad\left.-\frac{1}{18}\Lambda_5\left(\left(\Lambda_5 + \frac{1}{6}\lambda^2\kappa_5^4\right)+\frac{7}{324}\lambda^4\kappa_5^8\right)\,\frac{y^4}{4!} +
\cdots\right]\label{tay1}
\end{eqnarray}
\noindent { Note that obviously in the brane $g_{\theta\theta}(r,0)=r^2$}.
Defining $\psi(r)$ as the deviation from a Schwarzschild form for $H(r)$ \cite{Maartens,ccc,kanti,rs09,rs02,rs03,Gian} as $
H(r) = 1 - \frac{2GM}{c^2r} + \psi(r)$,
for a large black hole with horizon scale $r \gg \ell$ it follows from Eq.(\ref{potential1}) that
\begin{equation}\label{psi}
\psi(r) \approx -\frac{4GM\ell^2}{3c^2r^3}.
\end{equation}
\noindent The formula above, together with Eq.(\ref{potential1}), can be forthwith 
derived from the RS analysis concerning small gravitational fluctuations in terms of KK modes, where
a curved background can support a  bound state  of the higher-dimensional graviton \cite{Randall2,garriga}. Besides, the effect of the KK modes on the metric outside a specific matter distribution on the brane was  incorporated by \cite{garriga} in the form of the $1/r^3$ correction to the gravitational potential. Such corrections in the inverse-square law were experimentally shown in \cite{kapp}.

Now, given a general static spherically symmetric  
\bege
g_{\mu\nu}dx^{\mu}dx^{\nu} =-N(r)dt^2+A(r)dr^2 + r^2d\Omega^2,\label{cmetric}
\enge\noi 
 the Casadio-Fabbri-Mazzacurati 4D black hole solution was obtained in \cite{rs06,rs09}. The Schwarzschild 4D metric is obtained when 
$N(r) = \left(A(r)\right)^{-1}$ and $N(r) = 1 - \frac{2GM}{c^2r}$. Its unique extension into the bulk is a black string warped horizon,
with the central singularity extending all along the extra
dimension, and the bulk horizon singular \cite{rs06,rs09,hawk}. If the Schwarzschild metric on the
brane is demanded with a regular AdS horizon, there is no matter confinement on the brane: in this case matter percolates into the bulk \cite{ccc}.
The condition $N(r) = \left(A(r)\right)^{-1}$ holds in the 4D case, although the most general solution  is the Reissner-Nordstr\"om one \cite{rs06,rs09,GCGR}, related to the case II analyzed in which follows.
The case I below concerns the function $N(r) = 1 - \frac{2GM}{c^2r}$ like the Schwarzschild case, but this time $A(r)$ to be calculated.  Both cases are profoundly investigated, as well as their prominent applications.

\subsection{Casadio-Fabbri-Mazzacurati black string: Case I}
\label{3.1}
This case was analyzed by Casadio, Fabbri, and Mazzacurati in \cite{rs06,rs09}, regarding the 
4D black hole solution (\ref{cmetric}). They obtained a solution of the Einstein's equations distinguished from the Schwarzschild one, provided by the metric coefficients \beq N(r)&=& 1- \frac{2GM}{c^2r}\qquad\quad\text{and}\qquad
 A(r)=\frac{1-\frac{3GM}{2c^2r}}
{\left(1- \frac{2GM}{c^2r}\right)\left(1-\frac{GM}{2c^2r}(4\beta-1)\right)}\label{ar}\eeq to be considered in (\ref{cmetric}). The solution (\ref{ar}) depends
on just one parameter and for $M \rightarrow 0$ one recovers the
Minkowski vacuum.

The Casadio-Fabbri-Mazzacurati black string classical horizon, in the brane, is the solution of the algebraic equation $(A(r))^{-1}=0$.
In order to extract phenomenological information of numerical calculations, we first consider in this Subsection the case where $\beta = 5/4$. Indeed, our aim is to analyze the pure braneworld corrected effects on the variation of luminosity in quasars, composed by a black hole  which presents Hawking radiation in the brane equal to zero \cite{rs06,rs09}. It makes feasible our analysis on the variation of quasars luminosity, purely due to braneworld effects.  Hawking radiation in the context of black strings
was investigated, e.g., in \cite{Ahmed:2011fi, Casadioharms}.
Note that the metric above was also derived as a possible geometry outside a star on the brane \cite{maart01}. The corresponding Hawking temperature is calculated in \cite{rs06}.
In comparison with Schwarzschild black holes, the black hole provided by this solution is either hotter or colder, depending upon the sign of $(\beta - 1)$.

The extension of these solutions into the bulk has  prominent importance addressed in \cite{rs06}. For
the Schwarzschild case, the singularity on the brane extends
into the bulk and makes the AdS horizon singular.
Notwithstanding, according to the analysis illustrated by the graphics below, Eq.(\ref{tay})
asserts that the black string solutions might be regular for supermassive black holes.

Taking into account the metric in (\ref{cmetric}), the classical standard black hole radius is given by
--- supposing $r\neq \frac{3GM}{2c^2}$ --- two solutions of Schwarzschild type $R_S = \frac{2GM}{c^2}$, for our choice of the parameter $\beta = 5/4$, providing zero Hawking black hole temperature. { The assumption   $r\neq 3GM/c^2$ is quite natural: for this case the 4D Kretschmann
scalar $
K^{(I)} =  R_{\mu\nu\rho\sigma} R^{\mu\nu\rho\sigma}$ diverges for $r = 0$ and $r = \frac{3GM}{2c^2}$ (see the Appendix). Now, the Gauss equation is well known to relate the 5D and the 4D Riemann curvature tensor as \begin{equation}
{}^{(5)}R^\mu _ {\;\; \nu\rho\sigma} = R^\mu_{\;\; \nu\rho\sigma}
 -K^\mu _ {\;\; \rho}K_{\nu\sigma} + K^\mu _ {\;\; \sigma}K_{\nu\rho}.\label{gauss}\end{equation} 
By taking the junction conditions into account, where
consequently $K_{\mu\nu}=-\frac{1}{2}\kappa_5^2 \left(T_{\mu\nu}+ \frac{1}{3}
\left(\lambda-T\right)g_{\mu\nu} \right)$, for the vacuum case here considered is follows that $K_{\mu\nu}=-\frac{1}{6}\kappa_5^2 \lambda g_{\mu\nu}$. By inserting it in the Gauss equation (\ref{gauss}), it implies that the 5D  Kretschmann scalar $ {}^{(5)}K =  {}^{(5)}R_{\mu\nu\rho\sigma} {}^{(5)}R^{\mu\nu\rho\sigma}$ for the Casadio-Fabbri-Mazzacurati type I black string also diverges for $r = 0$ and $r = \frac{3GM}{2c^2}$: the terms involving the extrinsic curvature in (\ref{gauss}) above are not capable to cancel the divergence provided by the 4D  Kretschmann scalar, in the computation for  $ {}^{(5)}K$}.

 Using the same procedure as \cite{Maartens,ROLDAO/CARLAO}, one can use the metric coefficients (\ref{ar})  in Eq.(\ref{expandir}) and calculate the black string warped horizon. 
   As asserted, for instance in \cite{rs06,rs09}, this analysis can be attempted
either numerically or by Taylor expanding all
5D metric elements in powers of the extra
coordinate. In the graphics below, we explicit the value for the { {black string warped horizon}, provided by  $\sqrt{g_{\theta\theta}(R_S, y)}$, where $R_S$ is the Schwarzschild radius. Further, $\lambda=\Lambda = 1=\kappa_5$  hereupon [$M_\odot$ denotes the sun mass]:}
\begin{figure}[H]\begin{center}
\includegraphics[width=3.0in]{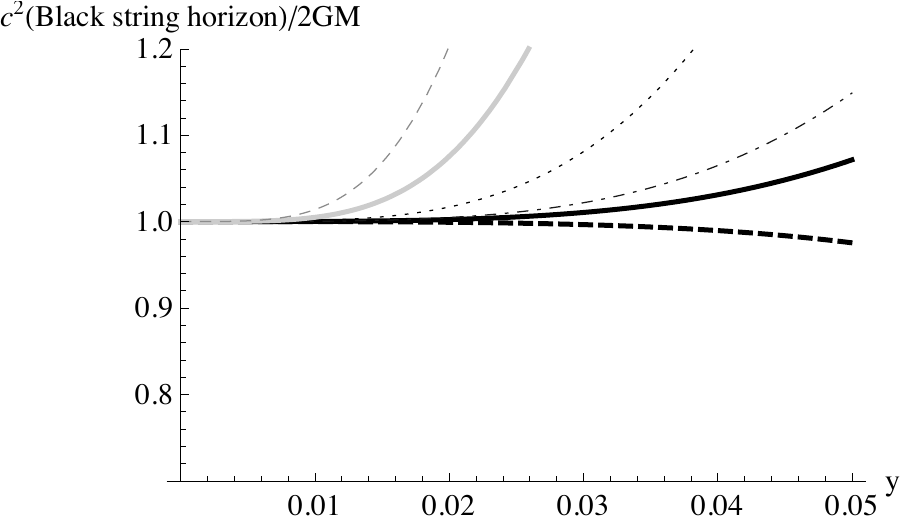}
\caption{\footnotesize\; Graphic of the brane effect-corrected   black string horizon $\sqrt{g_{\theta\theta}(R_S,y)}$ in the Casadio-Fabbri-Mazzacurati first solution, along the extra dimension $y$, for different values of the black hole mass $M$. For the  dash-dotted line $M= M_\odot$; for the black dashed line: $M = 10\,M_\odot$; for the  thick black line:  $M=10^2\,M_\odot$;  for the black dotted line:  $M=10^3\,M_\odot$; for the thick  gray line $M= 10^4\,M_\odot$;  for the  gray dotted line $M= 10^5\,M_\odot$. }
\end{center}\end{figure}

Fig. 1  evinces a very interesting profile for the black string horizon behavior along the extra dimension $y$ in gaussian coordinates. 
It indicates a critical mass $M$  (indeed our simulations provide $M \sim 73 M_\odot$) above which the associated black string warped horizon monotonically increases along the extra dimension. 
{ The black string is known to be placed in the bulk, in a tubular neighborhood
along the axis of symmetry. A singularity associated to the black string is a fixed point $y_0$ (fixed) in the axis of symmetry along the extra dimension, such that  the black string transversal slice has radius equal to zero. We show here that at the coordinate singularities
$r = 0$ and $r = \frac{3GM}{2c^2}$ there is a physical singularity for the black string at such values, 
irrespective of the value for $y$. In fact, the Kretschmann scalar $K =  {}^{(5)}R_{\mu\nu\rho\sigma} {}^{(5)}R^{\mu\nu\rho\sigma}$ diverges for such values (see the Appendix). Notwithstanding, 
 the black string warped horizon $\sqrt{g_{\theta\theta}(R_S,y)}$ does not equal to zero, as illustrated at the Fig. 1.  } 
\subsection{Casadio-Fabbri-Mazzacurati black string: Case II}
\label{3.2}
An alternative solution of  (\ref{cmetric}) is obtained in \cite{rs06,rs09} where the metric coefficients
 \begin{eqnarray}
N(r)&=& 1- \frac{2GM}{c^2r}+ \frac{2G^2M^2}{c^4r^2}(\beta-1),\qquad A(r)=\frac{1-3GM/2c^2r}
{\left(1- \frac{2GM}{c^2r}\right)\left(1-\frac{GM}{2c^2r}(4\beta-1)\right)}\label{ar111} 
\end{eqnarray} are considered in (\ref{cmetric}).
In order that the Hawking temperature be zero on the brane, the choice $\beta = 3/2$ is demanded  \cite{rs06}. The classical solution $R$ for the black hole horizon is given by $R = R_S$ and $R = 5R_S/2$, where $R_S$ denotes the Schwarzschild radius. It implies that the black string horizon now corrected by braneworld effects when (\ref{ar111}) is substituted in (\ref{expandir}), providing the graphic below.

\begin{figure}[H]\begin{center}
\includegraphics[width=3.0in]{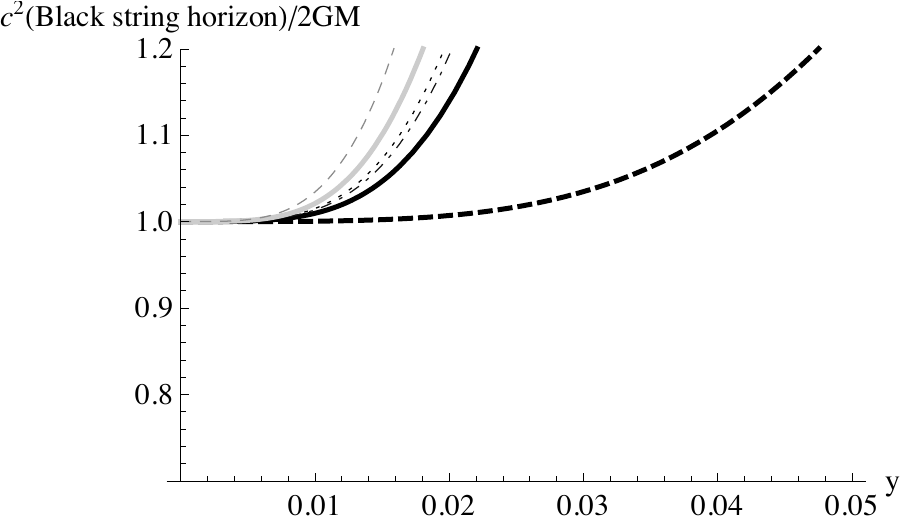}
\caption{\footnotesize\; Graphic of the brane effect-corrected Casadio-Fabbri-Mazzacurati  type II black string horizon $\sqrt{g_{\theta\theta}(R_S,y)}$, along the extra dimension $y$, for different values of the black hole mass $GM/c^2$ in the brane. For the  black  dashed line $M= M_\odot$; for the  gray line $M= 10\,M_\odot$; for the  black line:  $M=10^2\,M_\odot$;   the dash-dotted line:  $M=10^3\,M_\odot$; for the  dotted line: $M = 10^4\,M_\odot$; for the gray dashed line $M= 10^5\,M_\odot$; for the thick gray line $M= 10^6\,M_\odot$.  }
\end{center}\end{figure}

The  black string horizon profile along the extra dimension
is qualitatively similar for all values of $M$ { depicted here: the warped horizon always  increases monotonically.   
{ Furthermore,  under a similar analysis accomplished this time for the case II Casadio-Fabbri-Mazzacurati, and by taking into account the Kretschmann scalar (\ref{app2}) in the Appendix, we conclude that such expression diverges for 
$ r  = \frac{3GM}{2c^2}$, for $r=\frac{5GM}{2c^2}$ and $r = \frac{2GM}{c^2}$.  Contrary to the 
Schwarzschild metric, which presents the black hole horizon as a coordinate singularity --- which can circumvented by, e. g., the Kruskal-Szekeres coordinates --- and not as a physical singularity, 
the Kretschmann scalar for the Casadio-Fabbri-Mazzacurati case II metric indicates that each black hole horizon on the brane is a physical singularity, since it diverges for such values. Again, the terms involving the extrinsic curvature in (\ref{gauss}) above are not able to cancel the divergence induced by the 4D  Kretschmann scalar, when one calculates  $ {}^{(5)}K$. Hence, the black string also diverges for such values.}}

Fig. 2 indicates that  the Casadio-Fabbri-Mazzacurati (case II) black string horizon always increases.
Since the bulk has no fixed metric a priori, but 
it can be calculated from (\ref{tay}) taking into account the metric on the brane, we can calculate the bulk curvature using 
the metric coefficients in (\ref{tay}). 

Compact sources on the brane, such as
stars and black holes, have been investigated extensively. However, their description has proven
rather complicated and there is little hope to obtain analytic solutions. The present literature does in fact provide solutions on the brane \cite{Dadhich,c111,maart01,rs06}, 
perturbative results over the Randall-Sundrum background \cite{gidd,garriga}, and numerical treatments \cite{rs08}.
In \cite{c1}  the luminosity dissipation, the conditions for which a collapsing star generically evaporates and
approaches the Hawking behavior as the (apparent) horizon is formed, are also analyzed. 

\section{Corrections in the luminosity: braneworld effects}
\label{4}

 Once the black string behavior was previously analyzed along the extra dimension, we hereon aim to focus  on the corrections now restricted to the phenomena on the brane. These corrections are shown here to induce dramatic consequences on the quasars luminosity variation, due to the braneworld model considered. Due to its prominent importance on the analysis hereupon, the effect of higher dimensions in the gravity sector might begin to make their presence felt as the black hole horizon is approached. The case of braneworld black holes horizon corrections is explored hereon.

Quasars are astrophysical objects that can be found at large astronomical distances. Supermassive
stars and the process of gravitational collapse are showed to be able to probe  deviations from the 4D general
relativity \cite{ROLDAO/CARLAO}. 
The observation of quasars  in $X$-ray band can constrain the measure of the bulk curvature radius $\ell$.
Varied values for $\ell$ were used and tested, and no qualitative deviations have been detected. Table-top tests of Newton's law currently find no deviations down to the order $\ell \lesssim $ 0.1 mm. A more accurate magnitude limit improvement on the AdS$_5$ curvature $\ell$ is provided in \cite{emparan,Maartens} 
by analyzing the existence of stellar-mass black holes
on long time scales and of black hole $X$-ray binaries. Furthermore, the failure of current experiments using torsion pendulums and mechanical oscillators to observe departures from Newtonian gravity at small scales have set the upper limit of $\ell$ in the region $\ell\lesssim  0.2$ mm \cite{8888}. 

Regarding a static black hole being accreted, in a straightforward model the accretion efficiency $\eta$ is given by
\begin{equation}\label{eta}
\eta = \frac{GM}{6c^2R_S},
\end{equation}
\noindent
where $R_S$ denotes the black hole horizon, namely the black string horizon in the brane). 
The event horizon of the supermassive black hole is $10^{15}$ times bigger than the bulk curvature parameter $\ell$. This is not the case of mini black holes wherein the event horizon of magnitude orders smaller than $\ell$.
As proved in \cite{ROLDAO/CARLAO}, the solution above for the black string horizon can be also found in terms of the curvature radius $\ell$ \cite{nossoprd}. 
In the accretion rate model in \cite{shapiro},  
observational data for the luminosity $L$ estimates a value for $\ell$.
The luminosity $L$, due to the accretion in a black hole composing a quasar,
is a function of the bulk curvature radius parameter $\ell$, and  provided  by 
\begin{equation}\label{dl}
L(\ell) = \eta(\ell) \dot{M}c^2,
\end{equation}
\noindent
where $\dot{M}$ denotes the mass accretion rate. For a typical black hole of $10^{12} M_{\odot}$ in a supermassive quasar, the accretion rate 
is
$
\dot{M} \approx 2.1 \times 10^{19} {\rm kg}\, {\rm s}^{-1}
$ \cite{ROLDAO/CARLAO}.
Supposing that the quasar radiates in the Eddington limit \cite{shapiro} $
L = L_{{\rm Edd}} \sim 1.2 \times 10^{45}\left(\frac{M}{10^7 M_{\odot}}\right)\; {\rm erg\,  s^{-1}}$, 
 the luminosity is given by $L \sim 10^{47}\, {\rm erg\, s^{-1}}$.
From (\ref{eta}) and (\ref{dl}), the variation in the luminosity of a quasar composed by a supermassive black hole reads 
\begin{eqnarray}\label{dell}
\Delta L &=& \frac{GM}{6c^2}\left(R_{\rm brane}^{-1} - R^{-1}_S\right)\, \dot{M} c^2 = \frac{1}{12}\left(     \frac{R_S}{R_{\rm brane}}-1  \right)\dot{M}c^2,
\end{eqnarray}\noindent where $R_{\rm brane} = \sqrt{g_{\theta\theta}(r = R_S, y)}$ denotes the black string corrected horizon. 
In the next Subsection the variation of the quasars luminosity for the two Casadio-Fabbri-Mazzacurati black holes are depicted and analyzed. Furthermore, the difference in the luminosity between the pure Schwarzschild and the case of the solutions (\ref{ar}) and (\ref{ar111}) are computed and discussed in what follows. 

\subsection{Corrections in the quasar luminosity for the both Casadio-Fabbri-Mazzacurati solutions}
We want now to analyze how the corrections for the metric coefficients due to braneworld effects in \cite{Randall2,Maartens,nossoprd,ROLDAO/CARLAO} can affect the luminosity emitted by quasars composed by black holes
provided by the Casadio-Fabbri-Mazzacurati solutions (\ref{cmetric}), with coefficients (\ref{ar}, \ref{ar111}).
 The alteration in the black hole horizon definitely modifies the quasar luminosity. Its variation with respect to the pure  Schwarzschild luminosity is provided by Eq.(\ref{dell}) and here depicted, for the Casadio-Fabbri-Mazzacurati type I metric:
\begin{figure}[H]\begin{center}
\includegraphics[width=3.0in]{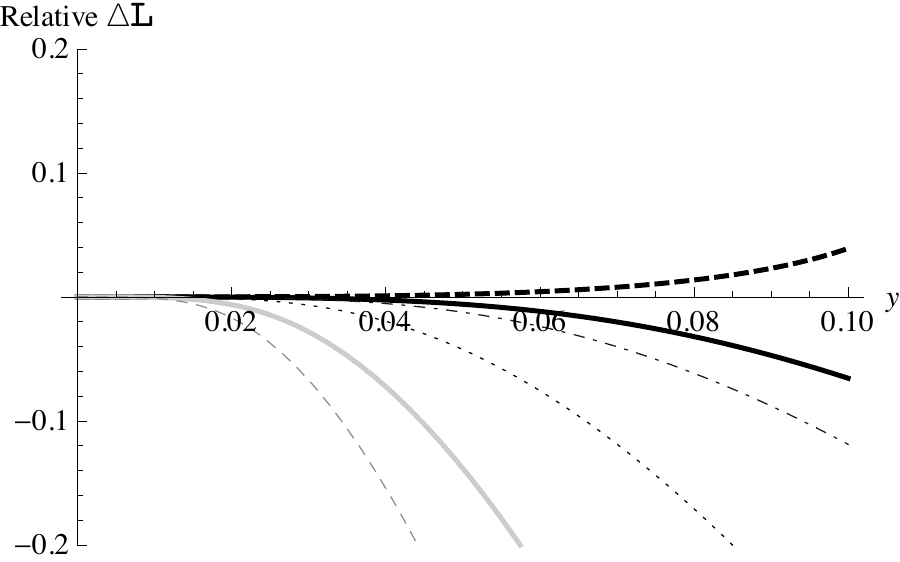}
\caption{\footnotesize\; Graphic of the relative variation of the luminosity $\Delta L/\dot{M}c^2$ in Casadio-Fabbri-Mazzacurati type I model as function of the black hole mass on the brane. For the dashed black line:  $M = 10^7 M_\odot$; for the continuous black line:  $M = 10^6 M_\odot$; for the dash-dotted line:  $M =  10^5 M_\odot$; for the dark dotted line:  $M = 10^4 M_\odot$; for the  light-gray thick line:   $M = 10^3 M_\odot$; for the gray dashed line  $M = 10^2 M_\odot$.}
\end{center}\end{figure}

Now the Casadio-Fabbri-Mazzacurati case II in Subsec. \ref{3.2} is analyzed, still adopting $\beta = 3/2$ in order to prevent Hawking radiation on the brane. It follows that similarly for the case I analyzed above,   the corrected black string horizon 
on the brane (namely the black hole horizon). One can show that for the corrections on the black string horizon, as seen from the brane, $\Delta L \sim 10^{30\pm 1}\;{\rm erg\, s^{-1}}$, for a typical supermassive black hole with $M \approx 10^9 M_\odot$ (case I and II respectively in the preceding Subsections \ref{3.1} and \ref{3.2}).  The figures  below regard respectively the  cases I and II of Casadio-Fabbri-Mazzacurati black strings horizon on the brane, analyzed in Sec. \ref{3}. In solar luminosity units $L_\odot \sim 3.9 \times 10^{33} {\rm erg\,s^{-1}}$, 
the variation of luminosity of a supermassive black hole quasar due to the correction of the horizon in a braneworld scenario is 
given by 
$
\Delta L \sim  10^{-3}\;L_\odot.
$
Naturally, this small but cognizable correction in the horizon of supermassive black holes 
implies a consequent correction in the quasar luminosity, via accretion mechanism. 
This correction is clearly regarded in the luminosity integrated in all wavelength. The detection of these corrections works in particular selected wavelengths, since quasars emit radiation
in the soft/hard $X$-ray band. We remark that the Schwarzschild braneworld corrected black string horizon on the brane was previously investigated in \cite{ROLDAO/CARLAO}, and 
in that case $\Delta L \sim  10^{-5}\;L_\odot.$ 

In addition,  the variation of the quasar luminosity regarding the Casadio-Fabbri-Mazzacurati type II model, with respect to the pure  Schwarzschild luminosity, is provided by Eq.(\ref{dell}) and here illustrated:
\begin{figure}[H]\begin{center}
\includegraphics[width=3.0in]{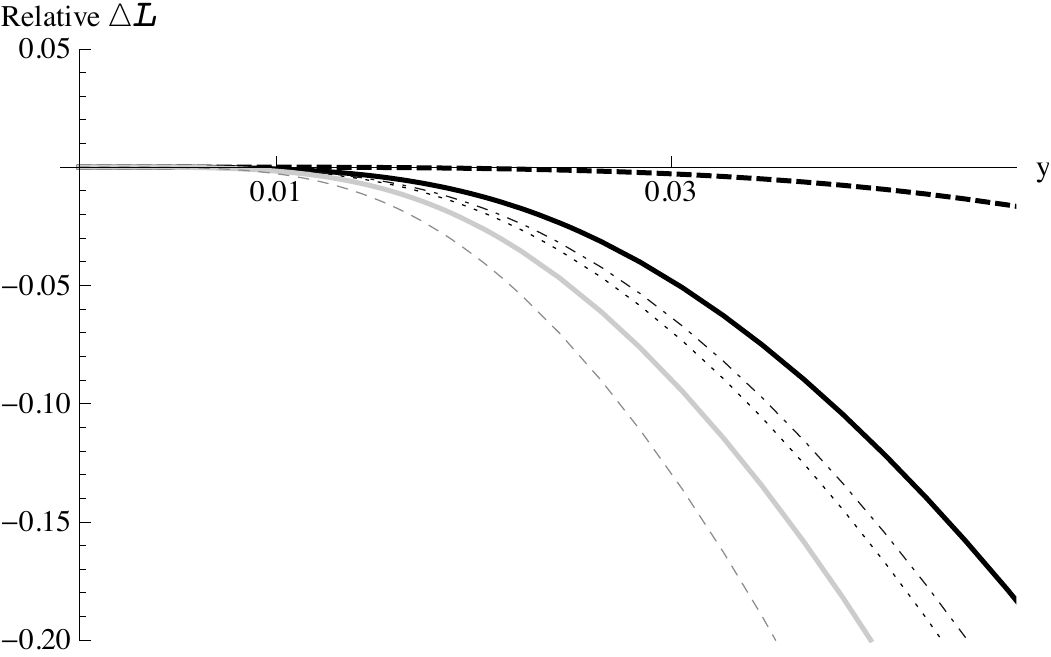}
\caption{\footnotesize\; Graphic of the relative variation of the luminosity $\Delta L/\dot{M}c^2$ in Casadio-Fabbri-Mazzacurati type II model as function of the black hole mass on the brane. For the  continuous black thin line:  $M = 10^6 M_\odot$; for the dash-dotted line:  $M =  10^5 M_\odot$; for the  light-gray thick line:   $M = 10^4 M_\odot$; for the gray dashed line  $M = 10^3 M_\odot$; for the dark-gray thick line:  $M = 10^2 M_\odot$.}
\end{center}\end{figure}

{ Figs. 3 and 4 evince the variation in the quasars luminosity in the 
Casadio-Fabbri-Mazzacurati types I and II metrics, respectively Eqs. (\ref{ar}, \ref{ar111}), with respect to the  Schwarzschild black hole luminosity.
The graphics reveal that the luminosity variation of the Casadio-Fabbri-Mazzacurati black holes, corrected by braneworld effects, is smaller compared to the Schwarzschild case. 
The exception is the curve in Figure 3 above the horizontal axis, which illustrates 
the general behavior of the Casadio-Fabbri-Mazzacurati type I black string  warped horizon, associated to a black hole mass $M \lesssim 73 M_\odot$. 
}
 

Delving into the analysis concerning the figures above, the general different profile between the Casadio-Fabbri-Mazzacurati black string warped horizon and the Schwarzschild horizon is expected. Their ratio(s) provides
the Figures 3 and 4 by Eq.(\ref{dell}) and the profile is encrypted in the underlying 
structure of Eq. (\ref{expandir}). Indeed,  the warped horizon is provided by $\sqrt{g_{\theta\theta}({R_S, y})}$ in (\ref{expandir}) when {\footnotesize{$\mu = \nu = \theta$}}. Besides, for the Schwarzschild metric the electric component of the Weyl tensor $ \mathcal{E}_{\theta\theta}$ equals zero, what do not happen to the Casadio-Fabbri-Mazzacurati metrics. Indeed, taking into account the metric (\ref{cmetric}), in  general  $
 {\cal E}_{\theta\theta} = -1 + \frac{1}{A} +\frac{r}{2A}\left(\frac{F'}{F} - \frac{A'}{A}\right)\neq 0$ for the coefficients in (\ref{ar}) and (\ref{ar111}).

{ As it is comprehensively discussed in \cite{sto1, sto2, max1, page}, the black hole may recoil away from the brane by the emission of Hawking radiation into the bulk, but not on the brane. We would like to emphasize that \emph{only} mini black holes in the Randall-Sundrum model are prevented to recoil away from the brane into the bulk \cite{sto2}. Notwithstanding, here the Randall-Sundrum model is not required and all  information about the bulk can be extracted from the Casadio-Fabbri-Mazzacurati metrics on the brane --- using (\ref{tay}) (and eventually further terms in $|y|^k$, for any $k$ positive, according to the required precision in the Taylor approximation). We considered terms up to $y^4$, since irrespective of the 
black hole horizon radius, the effective distance $y$ along the extra dimension equals 
the compactification radius \cite{Maartens}, as well as the effective size of the extra dimension probed by a graviton. 
Besides, our procedure considers suitable values for the post-Newtonian parameter
in order that the Hawking radiation on the brane is zero. 
The number of degrees of freedom of KK gravitons is much less than the number of standard model particles in the Hawking radiation in the bulk, and the black hole energy irradiated into the KK modes must be a small fraction of the total luminosity. Since the post-Newtonian parameter $\beta$
was chosen to prevent Hawking radiation in the brane, standard model particles on the brane with high enough energy --- larger than electroweak energy scale --- are capable to overcome the confining mechanism \cite{Casadioharms}. In this case the bulk standard model fields should be included among the KK modes. 
Such mechanism responsible for the possible black hole recoil from the brane corroborates to the astrophysical  phenomenology described in the figures above: the physical black hole radii $R_S = \sqrt{g_{\theta\theta}({R_S, 0})}$ are now \emph{effectively} dislocated into the bulk, and given by $R_{\rm brane} = \sqrt{g_{\theta\theta}({R_S, y})}$ in Eq.(\ref{dell}).}
Further discussion and details on the general behavior  encoded in the figures are presented in the next
Section.

\section{Concluding remarks and outlook}

Any phenomenologically successful theory in which our Universe is viewed as a brane must reproduce the large-scale predictions of general relativity on the brane. It implies that gravitational collapse of matter trapped on the brane provides the Casadio-Fabbri-Mazzacurati solutions on the brane: either a localized black hole or an extended black string solution, possessing a warped horizon. 
It is possible to intersect this solution with a vacuum domain wall and the induced metric is the ones presented in the analysis in  Subsections \ref{3.1} and \ref{3.2}.
In the case I,  our analysis is restricted to the case where $\beta = 5/4$ since for this values there is a zero (Hawking) temperature black hole associated \cite{rs06,rs09}. Since we want to extract physical information on the braneworld effects on the variation of luminosity
exclusively, we opted for this value for the parameter $\beta$, in such a way that the graphics, concerning this variation on the luminosity, take into account exclusively the braneworld effects, since Hawking radiation is shown to be suppressed with $\beta = 5/4$ for this metric. Analogously, for the case II the analysis is accomplished taking into account the value $\beta=3/2$ in Eq.(\ref{ar111}), as already discussed.

For the Casadio-Fabbri-Mazzacurati (types I and II) black holes 
the variation of luminosity of a supermassive black hole quasar due to the correction of the horizon in a braneworld scenario is 
given by 
$
\Delta L \sim  10^{-3}\;L_\odot.
$
On the other hand, the Schwarzschild braneworld corrected black string horizon on the brane was previously investigated in \cite{ROLDAO/CARLAO}, and 
in that case $\Delta L \sim  10^{-5}\;L_\odot$.  It shows that 
the Casadio-Fabbri-Mazzacurati black hole solutions, containing 
the post-Newtonian parameter, can probe two orders of magnitude more 
the variation in the quasars luminosity. 

Figs. 1 and 2 encode 
the brane effect-corrected  black string horizon, respectively  for the first and second black hole solution proposed by Casadio, Fabbri, and Mazzacurati, along the extra dimension $y$. The black string horizon behavior is obviously
different for distinct values for the black hole mass. For the second case, the warped horizon 
is always an increasing function of the extra dimension. For the first one, instead, it holds only for 
a black hole with mass $M \gtrsim 73 M_\odot$. Otherwise, the warped horizon  is a decreasing function along the extra dimension. 

Once the corrections related to the black string horizon behavior along the extra dimension are obtained, we focused 
on how such corrections can also alter the black string warped horizon. Our results are exposed and conflated in Figs. 3 and 4 illustrating the variation of luminosity of quasars --- supermassive black holes --- when the Hawking radiation in the brane is precluded, and the pure braneworld 
effect can be analyzed. The corrections of the luminosity regarding the Schwarzschild black string, along the extra dimension, can be probed by a black hole by recoil effects from the brane. The variation of the quasar luminosity is considerable in this case.

In \cite{nossoprd}  some properties of black holes were analyzed, in ADD \cite{todos} and Randall-Sundrum models. Mini black holes in ADD models have the first phase Hawking radiation  mostly in
the bulk and recoil effect to leave the brane. The analysis of the Casadio-Fabbri-Mazzacurati solutions in this paper sheds new light on 
mini black holes and their possible detection at LHC, since 
the preclusion of Hawking radiation can drastically modify the previous analysis about mini black holes in ADD and Randall-Sundrum braneworld models, as well as
the mini black holes radiation in LHC measurements. The method here introduced can be immediately applied in such context.

\section*{Acknowledgments}
R. da Rocha is grateful to Conselho Nacional de Desenvolvimento Cient\'{\i}fico e Tecnol\'ogico (CNPq)  476580/2010-2, 303027/2012-6, and
304862/2009-6 for financial support, and to Prof. J. M. Hoff da Silva for fruitful and valuable 
discussions and suggestions as well.  A. M. Kuerten thanks to CAPES for financial support. C. H. Coimbra-Ara\'ujo thanks Funda\c c\~ao
Arauc\'aria and Itaipu Binacional for financial support. 

\appendix
\section{Kretschmann scalars for Casadio-Fabbri-Mazzacurati metrics}
The 5D Kretschmann scalars associated to 
the black strings here discussed can be obtained by the
4D ones via Eq.(\ref{gauss}).

Since we aim to investigate the particular case where $\beta=5/4$ where 
there is no Hawking radiation, and pure braneworld effects can be 
probed, for this case the 4D Kretschmann
scalar $
K^{(I)} =  R_{\mu\nu\rho\sigma} R^{\mu\nu\rho\sigma}$ --- where the Riemann tensors used are the ones related to the Casadio-Fabbri-Mazzacurati case I --- is given by
 \begin{eqnarray}
K^{(I)} &=& \left(\frac{GM}{c^2r}\right)^4\left[\frac{(1-\frac{2GM}{c^2r})}{(1-\frac{3GM}{c^2r})}\right]^{2} \left\{\frac{GM}{c^2r}\left(3+\frac{7GM}{c^2r}\right)\right\}^{2}\nonumber
\\ &+& 4\left[\frac{8}{9}\left(1-\frac{2GM}{c^2r}\right)^{2} + \dfrac{G^2M^2}{2c^4r^{2}\left(1-\frac{3GM}{2c^2r}\right)^{2}}\left[\left(1-\frac{2GM}{c^2r}\right)\left(\dfrac{5}{2}-{\frac{3GM}{c^2r}}\right)\right]^{2}+\left(1-\dfrac{(1-\frac{2GM}{c^2r})^{2}}{(1-\frac{3GM}{2c^2r})}\right)^{2}\right] \nonumber\\
\label{app1}
\end{eqnarray} which diverges at $r=\frac{3GM}{2c^2}$ and  $r=0$. It agrees with the result in \cite{rs06, rs09} where
$K^{(I)} \propto (1-\frac{3GM}{2c^2r})^{-4}$  for values of $r$ near to the 
respective singularity $r=\frac{3GM}{2c^2}$. 
Now, for the Casadio-Fabbri-Mazzacurati case II metric (\ref{ar111}), the associated Kretschmann scalar, 
in the specific case here considered $\beta= 3/2$ the expression above is led to 
\begin{eqnarray}
K^{(II)} &=& \left(\frac{GM}{c^2r}\right)^{2}\left[\dfrac{\left(1-\frac{5GM}{2c^2r}\right)}{\left(1-\frac{3GM}{2c^2r}\right)\left(1-\frac{2GM}{c^2r}\right)}\right]^{2}\nonumber\\ & & \left[\left(\dfrac{3}{(1-\frac{3GM}{2c^2r})}-\dfrac{2}{(1-\frac{2GM}{c^2r})}-\dfrac{5}{2(1-\frac{5GM}{2c^2r})}\right)\frac{GM}{c^{2}r}\left(1-\frac{GM}{c^2r}\right)-\dfrac{\frac{2GM}{c^2r}}{\left(1-\frac{2GM}{c^2r}\right)}- \left(\frac{6GM}{c^2r}-1\right)\right]^{2}
\nonumber\\ &+& 4\left[\dfrac{8}{9}\left(1-\frac{2GM}{c^2r}\right)^{2}\left(1-\frac{5GM}{2c^2r}\right)^{2}{\left(1-\frac{GM}{c^2r}\right)^{-1}}\right.+\dfrac{(\frac{GM}{c^{2}r})^{2}}{2r^{4}(1-\frac{3GM}{2c^2r})^{2}}\left[2\left(1-{\frac{5GM}{2c^2r}}\right)\right.\nonumber\\&&\left.+\dfrac{5}{2}\left(1-\dfrac{2GM}{rc^2}\right)-\dfrac{3}{2}\left(1-\frac{2GM}{rc^2}\right)\left(1-\frac{5GM}{2rc^2}\right)\right]^{2}+\left. \dfrac{1}{r^{4}}\left(1-\dfrac{\left(1-\frac{2GM}{c^2r}\right)\left(1-\frac{5GM}{2c^2r}\right)}{\left(1-\frac{3GM}{2c^2r}\right)}\right)^{2}\right] 
\label{app2}
\end{eqnarray}

\end{document}